\documentclass[12pt]{article}
\usepackage{epsfig}
\usepackage{changebar}
\usepackage{amsmath}
\usepackage{amssymb}
\usepackage{amsbsy}
\begin{document}

\def\mezzo{\frac{1}{2}}
\def\intq#1{\int \frac{d^4#1}{(2\pi)^4}}
\def\intt#1{\int \frac{d^3#1}{(2\pi)^3}}

\author{R. Cenni\\
Istituto Nazionale di Fisica Nucleare -- sez. di Genova \\
Dipartimento di Fisica dell'Universit\`a di Genova \\
Via Dodecaneso 33 -- 16146 -- Genova -- Italy }
\title{On the $f$  Sum Rule and its Extensions}
 \maketitle
\begin{abstract}
The $f$ sum rule is derived in a non-relativistic frame and connected, via Ward
Identities, to the two-photon term of the Compton scattering. A
generalisation to isospin symmetry in the nuclear case 
 is discussed and linked to the 
Meson Exchange Currents. 
The extension to a fully relativistic theory is then discussed and it
is shown that the energy-weighted sum rule becomes a relation between the 
particle-hole and particle-antiparticle emission. Moreover the generalisation 
to isospin symmetry is derived and provides non-perturbative results.
\end{abstract}

\section{Introduction}
\label{sec:1}

In this paper we discuss the energy-weighted sum rule of the scalar-isoscalar
nuclear response ($f$) and its extension to a scalar-isovector one 
($f^\prime$) in a (as far as possible) non-perturbative frame.
  
As an introduction,  we examine the non-relativistic  sum rule.
This topic is well known since a long time (see, e.g., \cite{NoPi-66-B} 
for the very beginning and 
\cite{OrTr-91} for a comprehensive review of the subject), but still some 
new facets need to be explored. 

As a second step we connect the sum rule with the asymptotic 
behaviour of the polarisation propagator and, via Ward Identities (WI) 
with the two-photon term
in the electro-magnetic (e.m.) lagrangian (for an electron gas) or with a 
suitable generalisation of it in nuclear physics.

Next we consider fully relativistic schemes,
where  the previous derivation breaks down because of the absence of
two-photon terms, and
we got the seemingly paradoxical result that the sum rule 
is vanishing. This is not a paradox however, because now
the $f$-sum rule takes the form of a compensation between the nuclear response
in the space-like region and a reduction of the response
in the time-like region induced by the nuclear medium
(Pauli Blocking). The renormalisation procedure plays here a central role.

Finally this
approach opens the way to the
extension  of the sum rule
to the scalar-isovector channel ($f^\prime$ sum rule): 
while in a potential theory the sum rule
receives an extra contribution from the isospin dependence of the potential
(if any: when the potential is isospin-independent then $f$ and  $f^\prime$ 
coincide), when the potential is replaced by a dynamical meson exchange, 
then the above contribution  disappears but the sum rule is governed
by the averaged squared mesonic field.

\section{Generalities on the $f$ Sum Rule}
\label{sec:2}

To begin with, the $f$ sum rule, in  a non-relativistic frame,  reads
\begin{equation}
\Xi_1=\int\limits_{0}^{\infty}\omega d\omega\,R_{S=0}^{T=0}(q,\omega)
=\frac{q^2}{2m}N
\label{fpsr1}
\end{equation}
where $R_{S=0}^{T=0}$ is the nuclear response to a scalar-isoscalar probe 
and $N$ is the number of nucleons.
$R_{S=0}^{T=0}$ is linked to the corresponding 
polarisation propagator $\Pi_{S=0}^{T=0}$ by
\begin{eqnarray}
R_{S=0}^{T=0}(q,\omega)&=&-\frac{1}{\pi}\Im\int dx\int dy\, e^{iq\cdot (x-y)}
(-i)\frac{<{\mathbf \Psi_0}|T\left\{\rho(x),\rho(y)\right\}|{\mathbf
\Psi_0}>}{<{\mathbf \Psi_0}|{\mathbf
\Psi_0}>}\nonumber\\
&=&-\frac{1}{\pi}\Im\Pi_{S=0}^{T=0}(q,\omega)
\label{fpsr2}
\end{eqnarray}
and in turn $\Pi_{S=0}^{T=0}$, in Fourier transform and 
in Lehmann representation (LR), reads
\begin{eqnarray}
\lefteqn{\Pi_{S=0}^{T=0}(q,\omega)=\sum_{n}}
\label{sr19}\\
&&\frac{(E_n-E_0)\left\{\left|<{\mathbf \Psi_0}|\tilde\rho
(q)|{\mathbf \Psi_n}>\right|^2+\left|<{\mathbf \Psi_0}|\tilde\rho
(-q)|{\mathbf \Psi_n}>\right|^2\right\}}{\omega^2-(E_n-E_0)^2+i\eta}
\nonumber
\end{eqnarray}
($\tilde\rho(q)$ being the Fourier transform of $\rho(x)$: note that 
 $\tilde\rho(q)^\dagger =
\tilde\rho(-q)$; further, $\{|{\mathbf \Psi_n}>\}$ is the set of the
excited states of the system and $\{E_n\}$ the set of the 
corresponding energies). 
The LR   also shows that $\Pi_{S=0}^{T=0}$ is  even in $\omega$ and, at
fixed $\mathbf q$,
at a first sight seems to  behave asymptotically like $\omega^{-2}$.
This statement is by one side crucial, 
but is, on the other side, not immediately proven and deserves  more comments.

In a non-relativistic frame no renormalisation
is needed, i.e., no counter-term (that would dominate the asymptotic behaviour 
of $\Pi_{S=0}^{T=0}$) are required, 
but nevertheless our statement, that is equivalent to claim that the series
$$\sum_n(E_n-E_0)\left\{\left|<{\mathbf \Psi_0}|\tilde\rho
(\pm q)|{\mathbf \Psi_n}>\right|^2\right\}$$
is finite, can be easily
proven only in a perturbative frame.

Assume for sake of simplicity  the infinite nuclear matter limit:
there  $\Pi_{S=0}^{T=0}$, intended as function of the complex variable
$\omega$ at fixed ${\bf q}$ has a cut along the 
real positive axis; the cut at 
the lowest level (Free Fermi Gas, or FFG) ranges between
$\frac{q^2}{2 m}-\frac{q k_F}{m}$ (or 0) and $\frac{q^2}{2 m}+\frac{q k_F}{m}$
(the energy range allowed for a particle-hole pair [$p$-$h$; we shall denote 
particle-antiparticle pairs with $p$-$\overline p$]) and   thus has a 
finite upper bound;
this in turn amounts to say that the integral over the energy-weighted
spectral function of the polarisation propagator is also finite.
At higher orders the size of the response region increases, 
depending on the maximum number of allowed $p$-$h$ pairs
but still is bounded from above, so that the point $\omega
=\infty$ is regular  at each  order and the behaviour of  
$\Pi_{S=0}^{T=0}\sim\omega^{-2}$ is ensured. 

In the case of finite nuclei the proof is laborious, 
because the response region 
is extended up to $\omega=\infty$ but, again in a perturbative frame, it
falls down sufficiently quickly with $\omega$ and the correct asymptotic
behaviour is ensured again.

Since however the 
perturbative series  is (likely) asymptotic, we cannot rule out {\em a priori}
the existence of non-perturbative phenomena like instantons or so, that could
destroy the required behaviour $\Pi_{S=0}^{T=0}\sim\omega^{-2}$

With this proviso (even more relevant in the relativistic 
case, because  in the QCD vacuum instantons exist indeed) 
we come back to the sum rule, where a
direct calculation provides
\begin{equation}
\Xi_1(q)=\frac{1}{2}<{\mathbf \Psi_0}|\left[\tilde\rho
(-q),\left[H,\tilde\rho(q)\right]\right]|{\mathbf \Psi_0}>
\label{fpsr3}
\end{equation}
that, if $\rho$ commutes with the potential (i.e., if the potential is
local), can be written in the 
``canonical'' form (\ref{fpsr1}).

\begin{figure}
\centerline{
\epsfig{file=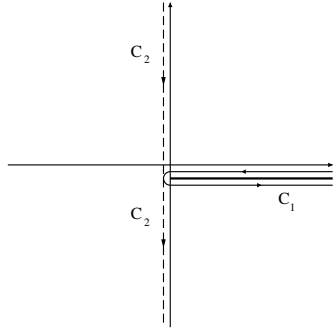,height=5cm}}
\caption{\protect\label{fig1}The integration paths for the $f$ sum 
rule}
\end{figure}

Let us now exploit the {\em assumed} analytical properties of 
$\Pi_{S=0}^{T=0}$ 
to link  its asymptotic behaviour to the sum rule . From the LR
 $\Pi_{S=0}^{T=0}$ is analytic on the 
whole first Riemann sheet but for a  cut along the real axis. 
Thus the sum rule can be rewritten as an integral over the whole
$\Pi_{S=0}^{T=0}$ as
\begin{equation}
\Xi_1=\frac{1}{2\pi i}\int_{C_1}\omega d\omega \Pi_{S=0}^{T=0}(q,\omega)\;,
\label{l1}
\end{equation}
the integration path $C_1$ being shown in fig. \ref{fig1}.
Would the integral be well behaved at the infinity, we could transform 
$C_1$ into a path along the imaginary axis, namely $C_2$.
Actually this is not allowed because the integrand behaves like 
$1/\omega$ at the infinity: we can however add and subtract  its 
asymptotic value by defining
\begin {equation}
\Pi^\infty (q) = \displaystyle\lim_{\omega \to \infty} \omega^2
\Pi_{S=0}^{T=0} (q, \omega)
\label{sr29}
\end {equation}
and then by splitting the integrand as 
\begin{equation}
\omega \Pi_{S=0}^{T=0} (q, \omega) = \frac{\omega \Pi^\infty (q)} 
{\omega^2 - a^2} + \left\{ \omega \Pi_{S=0}^{T=0}  (q, \omega) -
\frac{\omega \Pi^\infty (q)}{\omega^2 - a^2} \right\}
\label{sr30}
\end {equation}
where $a$ is an irrelevant  constant.
Now we can transform the integration path in the second 
term of (\ref{sr30}) and since its integrand is odd, its contribution 
vanishes. The only surviving part comes  from the first 
term of (\ref{sr30}) and  is
\begin{equation}
\Xi_1={1\over 2\pi i} \oint d\omega {\omega \Pi^\infty (q) 
\over
{\omega^2 - a^2}}= {1\over 2}\Pi^\infty(q)\;,
\label{fpsr5}
\end{equation}
the last integral being extended to a circle containing  $a$.
This relation is  fully general, i.e., it does not depend upon 
the  choice of $\rho$. Further, it holds in a relativistic 
context as well, provided the right behaviour of 
$\Pi_{S=0}^{T=0} (q, \omega)$ for $\omega\to\infty$ is ensured, and is
insensitive of a possible spin and/or isospin dependent interaction.

To exemplify let us consider the FFG. There the response function reads 
\begin{equation}
  R_{S=0}^{T=0}(q,\omega)=-\frac{4 V}{\pi}\Im\Pi^0(q,\omega)
\end{equation}
where $V$ is the box volume
and $\Pi^0$ denotes the Lindhard function without spin-isospin coefficients, 
that are factorized in front of it\cite{FeWa-71-B}.  The
simplest way to get the sum rule is to evaluate first the energy integral
\begin{align}
&\Xi_1^{FFG}(q) \label{ffg:can}\\
 =&4V\int \omega\,d\omega\,
 \int\frac{d^3k}{(2\pi)^3}\theta(|{\bf k}+{\bf q}|-k_F)
 \theta(k_F-k)\delta\left(\omega-\frac{\bf{q}^2}{2m}-\frac{{\bf q}\cdot{\bf k}}
 {m}\right)\nonumber
 \\
 =&4V\int\frac{d^3k}{(2\pi)^3}
 \theta(|{\bf k}+{\bf q}|-k_F)
 \theta(k_F-k)\left(\frac{{\bf q}^2}{2m}+\frac{{\bf q}\cdot{\bf k}}
 {m}\right)\nonumber
\end{align}
and a trivial calculation shows that the ``canonical value'' 
of the sum rule is reproduced.

To compare FFG with eq. (\ref{fpsr5}) we take the real part of
$\Pi^0$\cite{FeWa-71-B}, namely
\begin{eqnarray}
  \label{eq:repi}
 \lefteqn{\Re\Pi^0(q,\omega)=}\\
&&8\wp \int\frac{d^3k}{(2\pi)^3}\theta(|{\bf k}+{\bf q}|-k_F)
 \theta(k_F-k)
\frac{\strut\displaystyle\frac{{\bf q}^2}{2m}+\frac{{\bf q}\cdot{\bf k}}{m}}
{\strut\displaystyle\omega^2-
\left(\frac{\bf{q}^2}{2m}+\frac{{\bf q}\cdot{\bf k}}{m}\right)^2}
\nonumber
\end{eqnarray}
and since
\begin{equation}
 \lim_{\omega\to\infty}\omega^2 \Pi^0(q,\omega)=8
 \int\frac{d^3k}{(2\pi)^3}\theta(|{\bf k}+{\bf q}|-k_F)
 \theta(k_F-k) \left(\frac{q^2}{2m}+\frac{{\bf q}\cdot{\bf k}}{m}\right)\;:
\end{equation}
comparison with (\ref{ffg:can}) shows that
the canonical value is reached again.

\section{Sum Rules and Ward Identities}
\label{sec:3}

Now we consider the case of a conserved charge. We assume 
that   a symmetry group exists, that generates a conserved current 
$j^\mu(x)$, i.e., at a 
classical level, $\partial_\mu j^\mu(x)=0$ and, at a quantum level, 
no anomalies occur. In the following  $\rho(x)\equiv j^0(x)=j_0(x)$. 

The Ward Identities (WI) can be derived 
for any kind of Green's function.
In particular we need them  for the current-current 
polarisation propagator
\begin{equation}
i\Pi^{\mu\nu}(x-y)=
\frac{<{\mathbf \Psi_0}|T\left\{j^\mu(x),j^\nu(y)\right\}|{\mathbf 
\Psi_0}>}{<{\mathbf \Psi_0}|{\mathbf 
\Psi_0}>}\,.
\label{fpsr6}
\end{equation}
Our case   corresponds to a symmetry group $U(1)$
exploiting charge or baryon number 
conservation. The quantity denoted with
$\Pi_{S=0}^{T=0}$ in previous sections turns out to be $\Pi^{00}=\Pi_{00}$
(neglecting channel indices).

Before going on we need to better specify the meaning of (\ref{fpsr6}),
because it will play a central role in our derivation. Eq. (\ref{fpsr6})
is very general indeed and applies to any kind of systems, both in a
non-relativistic or relativistic frame.
In the latter case however a well known disease arises
when the ``naive'' definition of the $T$-product (this terminology
could be found for instance in  \cite{ItZu-80-B}; it could be better defined as
 Dyson's $T$-product, or simply $T_D$) is adopted, namely
\begin{equation}
  \label{naive}
T_D\left\{j^\mu(x),j^\nu(y)\right\}=j_\mu(x)j_\nu(y)\theta(x^0-y^0)
+x\leftrightarrow y\;:
\end{equation}
it has been discovered by Schwinger long time ago \cite{Sc-59} that in such
a case $\Pi^{\mu\nu}(x-y)$ (we shall denote with this symbol
the quantity defined in eq. (\ref{fpsr6}) when the $T$-product 
is specified to be the Dyson's one eq. (\ref{naive})) is not covariant.

It is customary to introduce another  $T$-product (also called Wick's
$T$-product, $T_W$) as
\begin{equation}
  \label{Wick}
T_W\left\{j^\mu(x),j^\nu(y)\right\}=T_D\left\{j^\mu(x),j^\nu(y)\right\}
+Sg^{\mu 0} g^{\nu 0}\delta(x-y)
\end{equation}
where the covariance is restored. The structure of the extra term in the above,
namely
$Sg^{\mu 0} g^{\nu 0}\delta(x-y)$, is well known (see for instance
\cite{Br-66} or \cite{ItZu-80-B}): we shall discuss it later, however, 
after having derived WI. We shall denote with $\tilde \Pi^{\mu\nu}(x-y)$
the current-current polarisation propagator derived from(\ref{fpsr6}) 
where the $T$-product has been replaced by $T_W$.

The ``caveat'' we must be aware of is instead that a Schwinger term
will alter the asymptotic (in $\omega$) properties of 
(the Fourier transform of) $\Pi^{\mu\nu}$. In fact the LR
eq.  (\ref{sr19}) applies to $\Pi^{\mu\nu}$ only; thus the Schwinger term
in $\tilde \Pi^{\mu\nu}$ will destroy the behaviour like $\omega^{-2}$.
A final remark is that covariant contact terms could still exist.
We shall see later in the relativistic infinite nuclear matter example
that a particular renormalisation scheme is able to get rid of them, 
thus preserving the asymptotic behaviour.

Coming back to our main job,  the derivation of the 
sum rule is a simple extension of the standard
way QFT uses to get WI (see for instance textbooks like
\cite{Am-78,ItZu-80-B}). The same results have already been obtained by
Takahashi, within a less powerful formalism\cite{Ta-86-B} and without linking 
it to the asymptotic behaviour of $\Pi^{00}$.

We write  the classical action of the system as
\begin{eqnarray}
\lefteqn{A=\int dx\, \psi^\dagger(x)\left\{i\frac{\partial~}{\partial x_0}
+\frac{\nabla^2}{2m}\right\}\psi(x)}
\label{fpsr7}\\
&&-\int dx\, dy\,\sum_{ij}\psi^\dagger(x){\cal O}_i\psi(x)
V_{ij}(x-y)\psi^\dagger(y){\cal O}_j\psi(y)
\nonumber
\end{eqnarray}
where the ${\cal O}_i$ are some spin and isospin (or, if case,  the 
identity) operators. 
The action is invariant under the global transformation
\begin{equation}
\psi\to e^{-i \Lambda}\psi\quad;\qquad\qquad\qquad
\psi^\dagger\to e^{i \Lambda}\psi^\dagger\;.
\label{fpsr9}
\end{equation}
Assuming $\Lambda\to\Lambda(x)$, up to the second order in $\Lambda$ one has
\begin{equation}
A[\Lambda]=A|_{\Lambda=0}+\int dx
j^\mu(x)\partial_\mu\Lambda(x)+\int dx 
B^{\mu\nu}(x)\partial_\mu\Lambda(x)\partial_\nu\Lambda(x)
\label{fpsr10}
\end{equation}
where
\begin{eqnarray}
\rho(x)=j^0(x)&=&\psi^\dagger(x)\psi(x)\\
{\bf j}(x)&=&-\frac{i}{2m}\left\{\psi^\dagger(x)\nabla\psi(x)-\left[
\nabla\psi^\dagger(x)\right]\psi(x)\right\}\\
B^{0\nu}= B^{\mu 0 }&=&0\label{bmn1}\\
B^{ij}&=&-\frac{1}{2m}\delta_{ij}\psi^\dagger(x)\psi(x)\;.
\label{bmn}
\end{eqnarray}
and spin-isospin dependence in the potential are irrelevant.
We now define a generating functional
\begin{eqnarray}
Z[{\cal A}_\mu]=\int D[\psi^\dagger,\psi]e^{iA+i\int j^\mu(x){\cal 
A}_\mu(x)dx}
\end{eqnarray}
where ${\cal A}_\mu$ is a classical external field, the link with
$\tilde\Pi^{\mu\nu}$ being
\begin{equation}
i\tilde\Pi^{\mu\nu}(x-y)=-\frac{\delta^2~~~~}{\delta{\cal A}_\mu(x)
\delta{\cal A}_\nu(y)}\log Z[{\cal A}_\rho]\Biggm|_{{\cal A}_\mu=0}\;.
\label{fpsr12}
\end{equation}
Note that here the covariant $T$-product (\ref{Wick}) comes into play. 
This is equivalent to say that the WI will take 
 a covariant form (as we shall check explicitly).
Observe also that the tie ordering implicit on the definition of the
path integral acts not only on the ordering of the currents but on the 
fields contained in it. Thus a further uncertainty could be added in the
covariant part of the contact term. This flaw will not trouble us however,
due to the renormalization mechanism we shall adopt in the relativistic case.

Coming to the derivation of the WI,  
we make the change of variable (\ref{fpsr9}) inside the path 
integral. By one side the integral remain unchanged, but on the other 
side $A$ transforms according to (\ref{fpsr10}) and $j^\mu(x)$ becomes
\begin{equation}
j^\mu(x)\to 
j^\mu(x)+2 B^{\mu\nu}(x)\partial_\nu\Lambda(x)\;.
\label{fpsr13}
\end{equation}
Since a change of the integration variable cannot alter $Z$ we can
equal the generating functional evaluated with $\Lambda=0$ with the one
where $A$ and $j^\mu$ are transformed according to (\ref{fpsr10}) and 
(\ref{fpsr13}). Next, by 
expanding in $\Lambda$ up to the first order, we get the identity
\begin{equation}
\int\hskip-2pt  D[\psi^\dagger,\psi]\int \hskip-2pt dx 
\left\{j^\mu(x)\partial_\mu\Lambda(x)
+2B^{\mu\nu}(x)\partial_\mu\Lambda(x){\cal A}_\nu(x)\right\}
e^{iA+i\int dx j^\mu(x){\cal A}_\mu(x)}=0
\label{ftsr14}
\end{equation}
Taking its functional derivative with respect to ${\cal A}_\mu(x)$, 
 putting  ${\cal A}_\mu(x)$ $=$ $0$ and further deriving with respect
to $\Lambda(y)$ we 
 get the WI we are interested in, namely
 \begin{equation}
   \label{eq:wi0}
   \int D[\psi^\dagger,\psi]\left\{\left(-i\partial_\mu^x j^\mu(x)
\right) j^\nu(y)-2\delta(x-y)\partial_\mu^x B^{\mu\nu}(x)\right\} e^{iA}=0\;,
 \end{equation}
where an extra term proportional to the 
density arises from the tensor 
$B^{\mu\nu}$. Using
(\ref{bmn}) and translating  (\ref{eq:wi0}) into expectation values of
physical quantities we  get the wanted W.I.:
\begin{equation}
\label{eq:addwi}
-i\partial_\mu^x<{\mathbf \Psi_0}|T\{j^\mu(x),j^\nu(y)\}|{\mathbf 
\Psi_0}>
=2\delta(x-y)<{\mathbf 
\Psi_0}|\partial_\mu^x B^{\mu\nu}(x)|{\mathbf \Psi_0}>
\end{equation}
 that in Fourier transform becomes
\begin{equation}
  q_\mu  \tilde \Pi^{\mu\nu}= 2 q_\mu <{\mathbf \Psi_0}|B^{\mu\nu}|{\mathbf 
\Psi_0}>
\end{equation}
namely a quite general result. Note that $B^{\mu\nu}$ is covariant
by construction, so also $\tilde \Pi^{\mu\nu}$ must be. 

In the  case at hand,
\begin{eqnarray}
q_\mu\tilde\Pi^{\mu 0}&=&0\\
q_\mu\tilde\Pi^{\mu i}&=&\frac{q^i}{m}\int d^3x<{\mathbf \Psi_0}|\rho
({\mathbf x})|{\mathbf 
\Psi_0}>=\frac{q^i}{m}N\;.
\end{eqnarray}
Combining these two relations and putting the $z$-axis along $\bf q$ we find
\begin{equation}
{1\over 2}q_0^2\tilde\Pi^{00}-{1\over 2}
|{\bf q}|^2\Pi^{33}=\frac{|{\bf q}|^2}{2m}N\;.
\label{fpsr15}
\end{equation}
(observe that $\tilde\Pi^{ij}=\Pi^{ij}$).
Finally we need to link $\tilde\Pi^{00}$ with
$\Pi^{00}$: we  get
\begin{equation}
{1\over 2}q_0^2\Pi^{00}+{1\over 2}q_0^2 <{\mathbf \Psi_0}|S|{\mathbf \Psi_0}>
-{1\over 2}|{\bf q}|^2\Pi^{33}=\frac{|{\bf q}|^2}{2m}N\;.
\label{fpsr15.01}
\end{equation}
We know from \cite{Br-66} that $S$ is vanishing in this case
(more precisely is vanishing as far as no squared time derivatives of the
fields are present in the lagrangian), so
in the limit $q_0\to\infty$ ($q_0\equiv \omega$)
the first term (being constructed in such a way to preserve
the asymptotic behaviour $\omega^{-2}$) provides 
${1\over 2}\Pi^\infty$, namely the sum rule, $|{\bf q}|^2\Pi^{33}$
is vanishing
because it  too  behaves like $\omega^{-2}$ (with the same provisos as for
$\Pi^{00}$) and the rhs 
term provides the ``canonical'' value of the sum rule.

Curiously, the above result does not need, in principle, any knowledge about 
$S$ but could be regarded as a way to derive it, because at the leading order
in $q_0^2$ (\ref{fpsr15.01}) just reads
\begin{equation}
{1\over 2}q_0^2  <{\mathbf \Psi_0}|S|{\mathbf \Psi_0}>=0\Longrightarrow  
<{\mathbf \Psi_0}|S|{\mathbf \Psi_0}>=0\;.  
\end{equation}

\section{A fully relativistic model}
\label{sec:5}

\subsection{The relativistic sum rule}
\label{sec:5.1}

A careful reader has surely recognised that the definition of $B^{\mu\nu}$
immediately leads in a relativistic dynamics
to get $0$ as the result of the sum rule, since in the Dirac theory
the nucleon current and the nucleonic $B^{\mu\nu}$ term read
\begin{equation}
  \label{eq:nuc}
  j_\mu^N=\overline\psi(x)\gamma_\mu\psi(x)\qquad\qquad B^{\mu\nu}=0\;.
\end{equation}

A brief comment about the current is needed. Here we assume nucleons
and mesons as structure-less, thus without form factors and anomalous momenta,
because both of them arise from the 
dressing of the elementary particles by means of the perturbative theory:
their explicit introduction would get rid of any non-perturbative results.
The same  holds {\em a fortiori} in the  case of
interacting nucleons and pions.

\begin{figure}[h]
  \begin{center}
    \epsfig{file=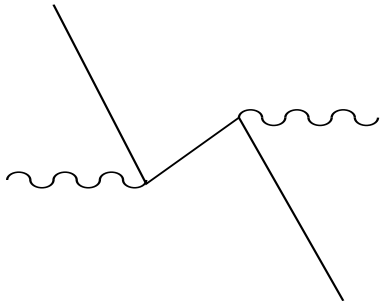,height=4cm,width=5cm}
    \caption{The relativistic version of the ``two-photon'' diagram.}
    \label{fig:3n}
  \end{center}
\end{figure}

Consider now a system of nucleons interacting
through an isoscalar meson and look to the $f$ sum rule.
In such a case no further contributions to $B^{\mu\nu}=0$ arise, and
the equivalent of eq. (\ref{fpsr15.01}) is
\begin{equation}
{1\over 2}q_0^2\Pi^{00}+{1\over 2}q_0^2  <{\mathbf \Psi_0}|S|{\mathbf \Psi_0}>
-{1\over 2}|{\bf q}|^2\Pi^{33}=0\;.
\label{fpsr15.1}
\end{equation}
that entails
\begin{equation}
  \label{eq:w}
  \Xi_1\equiv 0\;,\qquad\qquad  <{\mathbf \Psi_0}|S|{\mathbf \Psi_0}>=0\;.
\end{equation}
(again the old result about $S$ is recovered).
But, as contradictory as a vanishing sum rule could  seem, 
this results holds true, 
and stems  from the existence of the antiparticles \cite{LeRuOk-57}.
In fact intuitively the non-relativistic $B^{\mu\nu}$ can be seen as
the low energy limit of the ``two-photon'' time-ordered 
diagram of fig. \ref{fig:3n}.

The crucial point is that now the energy-weighted sum rule, as well as the 
Coulomb sum rule and the longitudinal response function itself are
non-vanishing even in the vacuum, provided the time-like region
is also considered.
It is in fact always possible to find a pair $q^0,{\bf q}$ such
that the probe can create a $p$-$\overline p$ pair. Even more,
the  sum rule for such a process is divergent.

Thus we are faced with two kind of problems: the first is that in deriving
the sum rule it is customary to extend the integration region up to
$+\infty$, thus going across the light cone and including 
nucleon-antinucleon pairs, the second is that the internal structure
of the nucleon cannot be simply factorized in front of the sum rules.

To both these question an answer, useful for practical calculation and
for a meaningful comparison with experimental data has been given in refs.
\cite{CeDoMo-97,AmCeDoMo-97,Ba-al-98}. In the conclusions we shall further
discuss  this topic

\subsection{The lowest-order case}
\label{sec:5.2}

The main difficulty met here is the separation of the nucleonic world
from the nuclear one and this requires an appropriate renormalisation
procedure. In order to understand the mechanisms leading to the unexpected
result (\ref{eq:w}) let us follow the FFG example we introduced
 in sect. \ref{sec:3}. 

The nucleon Green's function for a relativistic FFG (RFFG)
is usually written as
\begin{equation}
  \label{eq:gfm}
  S^{\rm m}(q)=\frac{\gamma\cdot q+m}{2E_q}\left[\frac{\theta(|{\bf q}|-k_F)}
  {q_0-E_q+i\eta}+\frac{\theta(k_F-|{\bf q}|)}{q_0-E_q-i\eta}
  -\frac{1}{q_0+E_q-i\eta}\right]\;,
\end{equation}
where $E_q=\sqrt{m^2+{\bf q}^2}$. The main difference with respect to FFG is 
of course the existence of a third term describing the propagation of
anti-nucleons. The evaluation of 
$\Pi^{\mu\nu}$ (remember that in this case $\Pi^{\mu\nu}$ and
$\tilde\Pi^{\mu\nu}$ coincide) reduces to
\begin{equation}
  \label{eq:pimnrel}
  \Pi^{\mu\nu}(q_0,|{\bf q}|)=-i\intq{k}j^\mu(q)S^{\rm m}(q+k)
j^\nu(q)S^{\rm m}(k)\;.
\end{equation}
The  explicit evaluation of the frequency integral, assuming
 $j^\mu=\overline{\psi}(x)\gamma^\mu\psi(x)$
leads to the cumbersome but on the other hand trivial expression 
\begin{eqnarray}
  \label{eq:pirel000}
\lefteqn{  \Pi^{00}(q_0,|{\bf q}|)=
 -2\intt{k}P(q,k)\Bigm|_{k_0=-E_k} }\\
&&\left\{ {\theta(|{\bf q+k}|-k_F)\theta(k_F-|{\bf k}|)\over
  q_0-E_{q+k}+E_k+i\eta}-
 {\theta(k_F-|{\bf q+k}|)\theta(|{\bf k}|-k_F)\over
  q_0-E_{q+k}+E_k-i\eta
  }\right\}
\nonumber\\
&&-2\intt{k}P(q,k)\Bigm|_{k_0=-E_k} {\theta(|{\bf q+k}|-k_F)
\over q_0-E_{q+k}-E_k-i\eta}
\nonumber\\  
 &&+2\intt{k}P(q,k)\Bigm|_{k_0=E_k}\Bigm|_{k_0=E_k}
{\theta(|{\bf k}|-k_F)\over q_0+E_{q+k}+E_k-i\eta}\;,
\nonumber
\end{eqnarray} 
where we have put, in order to simplify the notation,
\begin{equation}
  P(q,k)={m^2+2k_0(k_0+q_0)-(q\cdot k+k^2)\over
  E_k E_{q+k}}
\end{equation}
that is the term coming from the traces.
Note that the two last terms in (\ref{eq:pirel000}) are clearly divergent.
Note also that at the non-relativistic limit, $P\to 2$ and the first
two terms in (\ref{eq:pirel000}) are led back to the Lindhard function.

In order to remove the divergences, we introduce the polarisation 
propagator in the vacuum, namely
\begin{eqnarray}
 \label{eq:pirel001}
 \Pi^{00}(q)|_{\rm vacuum}&=&
 \Pi^{00}(q_0,|{\bf q}|)|_{k_F=0}\\
\nonumber
&=&-i\intq{k}j^\mu(q)S^0(q+k)
j^\nu(q)S^0(k)
\\
&=&-2\intt{k}P(q,k)\Bigm|_{k_0=-E_k}\frac{1}
{ q_0-E_{q+k}-E_k-i\eta}
\nonumber\\  
 &+&2\intt{k}P(q,k)\Bigm|_{k_0=E_k}
\frac{1}{q_0+E_{q+k}+E_k-i\eta}\;.\nonumber
\end{eqnarray}
where $S^0(q)=1/(\slash\hskip-6pt q-m+i\eta)$.
$\Pi^{00}(q)|_{\rm vacuum}$ is of course  divergent too and  requires 
a normalisation prescription not defined a priori. It describes
the propagation of a $p$-$\overline p$  in 
the vacuum and an obvious way of separating the medium effect from the
vacuum properties is that of subtracting (\ref{eq:pirel001}) 
from (\ref{eq:pirel000}), so getting
\begin{eqnarray}
\lefteqn{  
  \Pi^{00}(q_0,|{\bf q}|)|_{\rm reg}=
 -2\intt{k}P(q,k)\Bigm|_{k_0=-E_k}}
\label{eq:pirel002}
 \\
&&\times\left\{ {\theta(|{\bf q+k}|-k_F)\theta(k_F-|{\bf k}|)\over
  q_0-E_{q+k}+E_k+i\eta}-
 {\theta(k_F-|{\bf q+k}|)\theta(|{\bf k}|-k_F)\over
  q_0-E_{q+k}+E_k-i\eta
  }\right\}
\nonumber\\
&&+2\intt{k}P(q,k)\Bigm|_{k_0=-E_k} {\theta(k_F-|{\bf q+k}|)
\over q_0-E_{q+k}-E_k-i\eta}
\nonumber\\  
 &&-2\intt{k}P(q,k)\Bigm|_{k_0=E_k}
{\theta(k_F-|{\bf k}|)\over q_0+E_{q+k}+E_k-i\eta}
\nonumber
\end{eqnarray} 
that is now convergent as it should. Before going on let us remark that the 
prescription of subtracting from a diagram  its value taken at $k_F=0$ 
only works at the 
lowest order in perturbation theory. In more complicated case a rigorous
procedure exists and has been given in ref. \cite{AlCeMoSa-88} and roughly 
speaking amounts to subtract from any elementarily divergent sub-diagram
its value at  $k_F=0$ and then to remove the remaining superficial divergence
by subtracting the whole diagram again at  $k_F=0$. In other words, we can
use the Bogoljubov's recursion formula and replace each counter-term considered
there with the corresponding sub-diagram in the vacuum.

Note further that this procedure is quite independent by the regularisation
scheme we use (dimensional or Pauli-Villars regularisation or any else)
because the counter-terms so introduced pertain to the vacuum and have to 
subtracted too. So we only need to know that the theory is renormalisable
in the vacuum, but the results does not depend (being physically meaningful)
upon the chosen regularisation scheme.

$\Pi^{00}$ at large $q_0$ seems to  behave like $q_0^{-1}$.Since 
however the function must be even, due to its bosonic character,
then the correct behaviour as
$q_0^{-2}$ is ensured. Starting from the above equation and remembering that
the longitudinal response function for a FFG is connected to $\Pi^{00}$ by
$R_L=-{V\over \pi}{\rm Im}\Pi^{00}$,
 a cumbersome but conceptually simple calculation
provides the response function for a RFFG. Its  explicit form can be found 
in many papers, like, say, \cite{Do-al-92,CeDoMo-97,AmCeDoMo-97}.
In the last reference the response in the time-like region is also derived. 

For the present purpose the best choice is not to use the explicit form
of the response but, instead, to carry out first
the integral over $q_0$ of the imaginary part of (\ref{eq:pirel002}).

Let us observe first that
\begin{equation}
  -\frac{V}{\pi}\Im\Pi^{00}(q_0,|{\bf q}|
)|_{k_F=0}=2V\intt{k}P(q,k)\Bigm|_{k_0=-E_k}\delta(q_0-E_{q+k}-E_k)
\label{srv}
\end{equation}
and consequently
\begin{equation}
  -\frac{V}{\pi}\hskip-1pt\int\hskip-1pt q_0dq_0\;\Im\Pi^{00}(q_0,|{\bf q}|
)|_{k_F=0}=2V\hskip-1pt\intt{k}\hskip-1pt(E_{q+k}E_k)P(q,k)\Bigm|_{k_0=-E_k}
= \infty\;,
\end{equation}
i.e., in the vacuum the $p$-$\overline p$ contribution to the energy-weighted
sum rule is infinite. Thus when subtracting the vacuum we also subtract
this contribution. The meaning of the various  terms in (\ref{eq:pirel002})
is now clear: the first two terms give a positive response and describe
a $p$-$h$ pair creation (a nucleon is ejected from  the nucleus); 
the two remaining describe {\em
the correction to  the $p$-$\overline p$ creation due to the existence
of the Fermi Gas, that produces a Pauli blocking effect. Thus since the Pauli 
principle  inhibit the response
with respect to (\ref{srv}), when the vacuum is subtracted its 
contribution is negative.}

To exemplify, for $q^0>0$,
\begin{eqnarray}
  \label{eq:srrffg}
  \Im\Pi^{00}(q_0,|{\bf q}|)&=&2\pi\intt{k}P(q,k)\Bigm|_{k_0=-E_k} \\
&\times&\theta(|{\bf q+k}|-k_F)\theta(k_F-|{\bf k}|)\delta(q_0-E_{q+k}+E_k)
\nonumber\\
&-&2\pi\intt{k}P(q,k)\Bigm|_{k_0=-E_k} 
\nonumber\\
&\times&\theta(k_F-|{\bf q+k}|)
\delta(q_0-E_{q+k}-E_k)
\nonumber
\end{eqnarray}
and the integral over $q_0$ is trivial. We thus get
\begin{eqnarray}
  \Xi_1^{RFFG}&=&2V\intt{k}(E_{q+k}-E_k)P(q,k)\Bigm|_{k_0=-E_k,q_0=E_{q+k}-E_k}
\nonumber\\
&\times&\theta(|{\bf q+k}|-k_F)\theta(k_F-|{\bf k}|)\label{eq:ffgr2}\\
&-&2V\intt{k}(E_{q+k}+E_k)P(q,k)
\Bigm|_{k_0=-E_k,q_0=E_{q+k}+E_k}
\nonumber\\
&\times&\theta(k_F-|{\bf q+k}|)\;.\nonumber
\end{eqnarray}
By changing variable ($k\to-k-q$) in the second integral, with some simple 
algebra we get, for $q>2k_F$,
\begin{eqnarray}
  \label{eq:44bis}
  \Xi_1^{RFFG}&=&2V\intt{k}\theta(k_F-|{\bf k}|)\frac{1}{E_{q+k} E_k}
\nonumber\\
&\times&\biggl\{\left(E_{q+k}- E_k\right)\left[m^2+{\bf k}^2+{\bf k \cdot q}
+E_{q+k} E_k\right]
\nonumber\\
&+&\left(E_{q+k}+ E_k\right)\left[m^2+{\bf k}^2+{\bf k \cdot q}
-E_{q+k} E_k\right]\biggr\}
\\&=&4V\intt{k}\theta(k_F-|{\bf k}|)\frac{{\bf k \cdot q}}{E_k}=0\;,
\end{eqnarray}
and the cancellation between 
 $p$-$h$ and  $p$-$\overline p$ is complete so that
the sum rule vanishes indeed. We can also exactly evaluate the two 
contributions, getting
$${\left[(|{\bf q}|-k_F)^2+m^2\right]^{3\over 2} \left[{\bf q}^2+3|{\bf q}|
k_F+k_F^2+m^2\right]\over 20 k_F^3 |{\bf q}|}N$$ 
for the $p$-$h$ term and the opposite for the $p$-$\overline p$ one.
The above expression is
cumbersome and not enlightening. Two limits are 
more interesting: for small ${\bf q}$ we get
$${{\bf q}^2\over 2\sqrt{k_F^2+m^2}} N\;;$$
if we remember that in the non relativistic limit $k_F$ is negligible too when 
compared with $m$, for small $k_F$ (a quite reasonable limit, as said 
before) and without constraint over ${\bf q}$ we also get
$${ {\bf q}^2\over 2\sqrt{{\bf q}^2+m^2}} N\;.$$
Needless to say the $p$-$\overline p$ contributions provide the same 
result up to a sign.

Now the physics is clear: the sum rule taken by itself (i.e. with no 
subtraction of the vacuum) is divergent. When the vacuum is subtracted
(and hence the theory is renormalised) we also subtract the possibility 
of creating $p$-$\overline p$ pairs in the vacuum.
What the sum rule tells us  is that the integrated energy-weighted strength
of the $p$-$h$ creation is just equal to the reduction of 
$p$-$\overline p$ creation due to the presence of a nuclear
 medium that imposes,
through the Pauli principle, that the emitted particle must have a 
momentum larger than $k_F$ (Pauli blocking).
Here we have explicitly evaluated $\Xi_1^{RFFG}$ because it clarifies
a result that at a first sight could seem paradoxical. Of course at each 
perturbative order we still get $\Xi_1=0$, since this outcome was derived
in full generality  via WI.

Thus the sum rule maintains its validity in spite of the seemingly 
contradictory result. The key ingredient is the separation of the vacuum:
this is something more than a renormalisation procedure: it also ensures,
in fact, that the regularised part of $\Pi^{00}$, as well as of
$\Pi^{33}$, conserves a behaviour
 $\sim q_0^{-2}$ at the infinity. 

As a final comment, we want to outline the analogy between the effect
described in this subsection and the creation of $e^+\,e^-$ pairs in
the electric field originated by a medium-heavy nucleus \cite{GoLvPe-83}
or the so-called Darmstadt effect, where again an $e^+\,e^-$ pair is created
by heavy ions collision \cite{Ki-86}.

\section{A Sum Rule for the Isovector Channel}
\label{sec:6}

The techniques developed so far  can also  be applied to the 
derivation of the $f^\prime$ sum rule, i.e., the integrated response to an 
isospin probe.

Following the path of the previous sections, we introduce
an isovector density of the form  $\rho^i(x)=\psi^\dagger\frac{\tau^i}{2}\psi$ 
and we get
\begin{equation}
\Xi_1^{T=1}(q)=\frac{1}{2}<{\mathbf \Psi_0}|\left[\tilde\rho^i
(-q),\left[H,\tilde\rho^i(q)\right]\right]|{\mathbf \Psi_0}>\;.
\label{fpsr3.1}
\end{equation}
Clearly if $H=T+V$ and if the potential 
is isospin-independent, then $[V,\tilde\rho^i(q)]=0$ and the
sum rule comes back to the ``canonical'' value (\ref{fpsr1}). If instead $V$
depends upon isospin, the sum rule is 
easily got by directly evaluating the commutator but is now
model-dependent.

We wish to consider, instead, a model where the 
Meson Exchange Currents (MEC) are explicitly embodied in the Lagrangian.

\begin{figure}[ht]
\centerline{
\epsfig{file=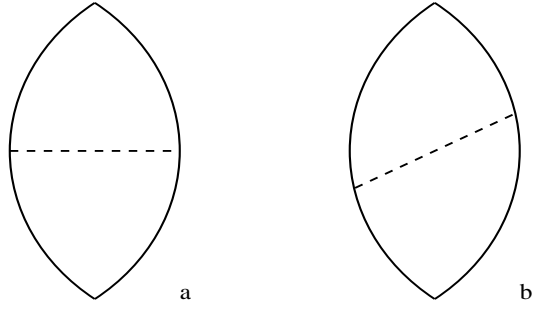,height=4cm,width=7cm}
}
\caption{The next-to-leading order corrections to the sum rule in a potential 
model (a) and in a theory with explicit meson exchange (b).These diagrams 
have to be intended as Goldstone diagrams}
\label{fig:4x}
\end{figure}

To understand the qualitative changes brought in by the meson dynamics 
consider the Goldstone diagrams of fig. \ref{fig:4x}.
Let us start from the diagram $a$: it represents
the next-to-leading order correction in a potential theory 
(the horizontal dashed line represents an instantaneous interaction)
and clearly it displays two energy denominators carrying the external incoming 
energy $\omega=q_0$. Thus its limit when multiplied by $\omega^2$ is 
finite.
If we however consider the diagram $b$, where the dashed line is inclined
to remind that a  meson is exchanged with its delay effects 
(or its intrinsic energy dependence) then the corresponding Goldstone diagrams
contains three energy denominators, each one carrying 
one $\omega$ and its behaviour is now like $\omega^{-3}$ 
and gives no contribution to
the sum rule. The same occurs for the 
energy-weighted sum rule for the spectral function and will be
explained in more details in a forthcoming paper. 

This simple analysis shows however that a dynamics for the exchanged mesons
is required because it intrinsically alter the structure of the $f^\prime$
sum rule. Let us in fact 
consider the following Lagrangian:
\begin{eqnarray}
\lefteqn{{\cal L}=\psi^\dagger(x)\left\{i\frac{\partial~}{\partial x_0}
+\frac{\nabla^2}{2m}\right\}\psi(x)}
\label{mec}\\
&&+{1\over 2}[\partial_\mu {\vec
\phi}(x)]^2-{\mu^2\over 2}{\vec \phi}^2(x)+i\frac{f_\pi}{\mu}\psi
^\dagger(x){\mathbf \sigma}\cdot {\mathbf \nabla}
\left({\vec \tau}\cdot {\vec \phi}\right)\psi(x)
\nonumber
\end{eqnarray}
The field $\vec\phi$ being an isovector meson,
three  conserved currents exist, with 0-components
\begin{equation}
  \label{eq:az1}
  j_0^i(x)\equiv\rho^i(x)=\psi^\dagger(x)\frac{\tau^i}{2}\psi(x)
  +\epsilon^{ijk}\phi^j(x)\partial_0\phi^k(x)
  \equiv \rho^i_N(x)+\rho^i_\pi(x)
\end{equation}
the second term representing a necessary feature of the isospin charge.
Physically it corresponds (for the $i=3$ component)
 to the pion-in-flight term in the isovector
part of the e.m. current. This term is often neglected in
nuclear calculations (while usually the 3-vector part is fully accounted for).

Denoting with $H_{\rm int}$ the $\pi N$ interaction term of the model,
an explicit calculation of the double commutator eq. (\ref{fpsr3})
provides the contribution (we assume for simplicity that $|{\mathbf \Psi_0}>$
is isoscalar)
\begin{equation}
\label{fpsr3.2}
-\frac{i}{3}\frac{f}{\mu}
\int d^3x <{\mathbf \Psi_0}|\psi^\dagger(x)\mathbf\sigma\cdot\mathbf\nabla
\vec\tau\cdot\vec\phi(x)\psi(x)|{\mathbf \Psi_0}>\delta_{ij}\;.
\end{equation}
However, the pion current 
term also exist and thus other three terms need to be considered, with one or
two nucleon currents replaced by the pionic ones. A simple (even if tedious)
calculation show that the three remaining terms have the same structure 
and the same coefficient as (\ref{fpsr3.2}), up to a sign, and 
all the four terms cancel out, giving
 the  result
\begin{equation}
\frac{1}{2}<{\mathbf \Psi_0}|\left[\tilde\rho^i
(-q),\left[H_{\rm int},\tilde\rho^j(q)\right]\right]|{\mathbf \Psi_0}>=0\;,
\label{fpsr3.3}
\end{equation}
and the sum rule regains the canonical value.
This result was expected (at least formally) both because  
 no $B^{\mu\nu}$ term is 
associated to the interaction Hamiltonians (that contains only one derivative),
and thus, following our previous derivation  no change was expected in the 
sum rule, and because the diagrams of fig. \ref{fig:4x} already suggested us 
that the energy dependence of the meson exchange should kill further
contributions coming from the interaction.

\begin{figure}
\centerline{
\epsfig{file=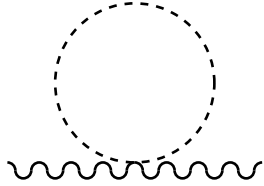}
}
\caption{\protect\label{fig2}The pion tadpole}
\end{figure}

The term $1/2 (\partial \phi^i_\mu)^2$ introduces a new feature in the model,
that is absent when MEC are handled statically, namely the arising of a term 
   $B^{\mu\nu}$ coming
from the free pion Lagrangian. It corresponds to the tadpole displayed in fig. 
\ref{fig2}, and clearly will affect, via WI, 
the sum rule. The present model is however neither covariant not renormalisable
so that the polarisation propagator is meaningless.
Thus in order to handle objects whose existence is ensured at any order 
let us  replace the model (\ref{mec}) with a fully covariant one, namely
\begin{equation}
  \label{eq:lag}
  A=\int d^4x\, \overline\psi(i\slash\hskip-6pt\partial-m)\psi+{1\over 2}
  (\partial_\mu\vec\phi)-{\mu^2\over 2}{\vec\phi}^2
  +ig \overline\psi \vec\tau\gamma_5\psi\vec\phi\;,
\end{equation}
where $\psi$ is an isospin doublet.

We follow now the same path of sect. \ref{sec:3}, replacing however
(with obvious meaning of the symbols) eq. (\ref{fpsr10}) with
\begin{equation}
A[\Lambda]=A|_{\Lambda=0}+\int dx
j^{i\mu(x)}\partial_\mu\Lambda_i(x)+\int dx 
B^{ij\mu\nu}(x)\partial_\mu\Lambda_i(x)\partial_\nu\Lambda_j(x)
\label{fpsr10iso}
\end{equation}
The associated isospin currents read
\begin{equation}
  \label{eq:cu}
  j_\mu^i=\overline\psi \frac{\tau^i}{2}\gamma_\mu\psi+\epsilon^{ijk}
\phi^j(x)\partial_\mu\phi^k(x)
\end{equation}
and the associated ``two-gauge boson term'' is given by
\begin{equation}
{B^{ij\mu\nu}}(x)=\frac{1}{2}
g^{\mu\nu}\left\{\frac{2}{3}{\vec\phi}^2(x)\delta^{ij}
-\left[\phi^i(x)\phi^j(x)-\frac{1}{3}{\vec\phi}^2(x)\delta^{ij}\right]
\right\}
\label{bmnp}\;,
\end{equation}
corresponding to the tadpole of fig. \ref{fig2}, and clearly  affects, via WI, 
the sum rule. 

Further, in the non-abelian case eq. (\ref{eq:addwi}) does not follow 
directly from (\ref{eq:wi0}) since now, in general, $[j_0^i,j_\nu^j]$
is not vanishing but is instead $\sim \epsilon{ijk}j_\nu^k$. In the simple 
case of isosinglet ground state the average value of this contribution
vanishes however. In this simplified situation also the isotensor term
of $B^{ij}{\mu\nu}$ is immaterial and we get for the WI the simplified 
expression
\begin{equation}
  \label{eq:wata}
  q^\mu\tilde\Pi_{\mu\nu}^{ij}=\frac{2}{3}q_\nu \delta^{ij}\int d^3x\, 
<{\mathbf \Psi_0}|\left[\vec\phi(x)\right]^2|{\mathbf \Psi_0}>
\end{equation}
where the last line holds provided the ground state is an isosinglet.
With the same assumption on the ground state the isospin structure of 
$\tilde\Pi_{\mu\nu}^{ij}$ also simplifies to $\tilde\Pi_{\mu\nu}\delta^{ij}$.
Thus, neglecting isospin indices and
following the same procedure as above, i.e., rewriting $\tilde\Pi^{00}$
in terms of $\Pi^{00}$ we end up with
\begin{equation}
  \label{eq:relpion}
  q_0^2\Pi_{00}+q_0^2 <{\mathbf \Psi_0}|S|{\mathbf \Psi_0}>
-{\bf q}^2\Pi_{33}=(q_0^2+{\bf q}^2)\frac{2}{3}
\int d^3x <{\mathbf \Psi_0}|\left[\vec\phi(x)\right]^2|{\mathbf \Psi_0}>\;.
\end{equation}
This result does not contain anymore the ``canonical value'' because of the 
cancellation described in the previous section but contains $S$.
Taking the leading term in $q_0^2$ in (\ref{eq:relpion}) we can again
derive the expression for it, namely
\begin{equation}
  \label{eq:cons}
   <{\mathbf \Psi_0}|S|{\mathbf \Psi_0}>
=\frac{2}{3}\int d^3x\, <{\mathbf \Psi_0}|\left[\phi^i(x)\right]^2|{\mathbf 
\Psi_0}>\;.
\end{equation}
Of course $S$, being a true Schwinger term,
 could also be written as a commutator (as done for instance 
in \cite{Br-66} and \cite{ItZu-80-B} chapt. 5.1.7). Note also that
$S$ is non-vanishing because of the term proportional to $(\dot\phi)^2$
in the lagrangian. It could be shown that, because of the existence of this
term, the translation from the lagrangian to the hamiltonian formalism is
not trivial and affects in particular $B^{\mu\nu}$ that is  altered
just by the amount $S$.

Note further that a subtraction of the vacuum is needed also for the Schwinger 
term,
as well as for the rhs of eq. (\ref{eq:cons}), making its expectation
value  finite in the medium.

Having determined $<{\mathbf \Psi_0}|S|{\mathbf \Psi_0}>$, 
we can cancel the highest order terms in $q_0^2$ in (\ref{eq:relpion})
and extend the energy-weighted sum rule to the isospin currents,
with the nontrivial result
\begin{equation}
  \label{eq:isov}
  \Xi^i={1\over 2} \lim_{q_0\to\infty} q_0^2 \Pi_{00}
  =\frac{1}{3}
{\bf q}^2\int d^4x\, <{\mathbf \Psi_0}|\left[\vec\phi(x)\right]^2|{\mathbf 
\Psi_0}>=\frac{{\bf q}^2}{2} <{\mathbf \Psi_0}|S|{\mathbf \Psi_0}>\;.
\end{equation}
Of course one still needs to evaluate diagrammatically the last term,
but we can also derive a non-perturbative result, namely
that
\begin{equation}
  \label{eq:finx}
  \Xi^i({\bf q}) = {\rm const} \times {\bf q}^2\;.
\end{equation}
or, in other word, the constant is model-dependent but the functional
dependence of the sum rule upon $ {\bf q}$ is fixed non-perturbatively
to be ${\bf q}^2$.
\section{Conclusions}
\label{sec:7}

In conclusion we have obtained the following results:

\begin{enumerate}

\item We have re-derived the $f$ sum rule in such a way to connect it with
the behaviour at infinity (in $\omega$) of the longitudinal polarisation
propagator and then, via WI, to the ``two-photon'' term 
$B_{\mu\nu}$ of the photon scattering amplitude.

\item The $f$ sum rule, when extended relativistically 
vanishes due to the cancellation between $p$-$h$ and $p$-$\overline p$
contributions. The first term is responsible of the non-relativistic 
sum rule (and in fact the relativistic Free Fermi Gas calculation reproduces 
the limiting case for small $q$) while the second term completely 
cancel the former
because the $p$-$\overline p$ is inhibited by the Pauli blocking.

\item In the case of $f^\prime$ sum rule a potential 
theory leads to the violation of the ``canonical value''
$$\Xi_1=\frac{{\bf q}^2}{2m}A$$
but if we include the mesonic field in the lagrangian a further current 
is added and when the missing part of the current is accounted for, then
we find again
$$<{\mathbf \Psi_0}|\left[\tilde\rho^i
(-q),\left[H_{\rm int},\tilde\rho^j(q)\right]\right]|{\mathbf \Psi_0}>=0\;.$$

\item In the $f^\prime$ case however we need a fully relativistic 
theory. Here the non-perturbative conclusion we can draw is that
\begin{equation}
  \label{eq:last}
  \Xi^i({\bf q})= S {\bf q}^2\;,
\end{equation}
where $S$ is the Schwinger term
(suitably renormalised) that is given by
\begin{equation}
  \label{eq:sch1}
  S=\frac{2}{3}\int d^3x\;<{\mathbf \Phi_0}|
\left[\vec\phi(x)\right]^2|{\mathbf \Phi_0}>\;.
\end{equation}
Note that the Schwinger term arises when one requires the Lorentz covariance
and thus is well defined only in a fully covariant model. This was 
not the case for the model (\ref{mec}) because there the lagrangian was not
covariant and the definition of the Schwinger term was meaningless.
\end{enumerate}

These conclusions still open new perspectives: 
\begin{enumerate}
\item First of all it will be interesting, as previously mentioned, 
to reproduce these results in a Goldstone expansion scheme, that will clarify
the rather formal aspect of the present paper.

\item New possibilities are opened, for instance the study of the sum rule in 
a parity violating response.

\item Finally, it seems to be exciting to analyse a symmetry group like
$SU(2)\otimes SU(2)$ where the Adler anomaly breaks the symmetry at the first 
order in the loop expansion.
\end{enumerate}

Before concluding, a last topic should be mentioned. The present paper makes
an attempt to draw non-perturbative conclusions for the $f$ and $f^\prime$
sum rules. In so doing we have discovered that this could be done 
only extending the integrations beyond the light cone (by the way, the same
limitation affected also ref. \cite{Wa-83}). Of course in comparing with 
experimental data the above is a serious disease. One could think maybe to
$\overline{p}$ production in heavy ions collisions, in analogy with the 
Darmstadt effect, but we dot believe that data coming from so different frames
could be seriously compared. Thus experimentally one is forcedly limited by 
the light cone. A series of papers \cite{CeDoMo-97,AmCeDoMo-97,Ba-al-98}
attempted to substantiate a set of sum rules in the space-like region.
The algorithm proved to be stable with respect to different nuclear models
but still was limited by the initial PWIA assumption.
A further disease overcome by those paper was the interconnection, in a
relativistic frame, of the nuclear and nucleon dynamics, that manifests
itself in a dependence upon $k_F$ of the form factors: the counterpart in the
present paper is that the renormalization procedure, and hence the dressing
of the vertices, i.e. the introduction of the form factors, must be
carried out contextually with the perturbative expansion. 

A major point to be understood in this frame is how the cancellations
between $p-h$ and $p-\overline{p}$ realize themselves.
To reach this goal a perturbative analysis is required: it will show
how, when some degrees of freedom are frozen,  a relevant 
contribution to the sum rule comes from the tail: in fact in the above 
situation we implicitly fix the infinity below the threshold of the
frozen degrees of freedom, and this in turn implies an Ansatz on
the tail of the integrand in the sum rule. From our analysis this fact
further implies an overestimate of the sum rule. 

The above discussion seems to be rather irrelevant for the $f$ sum rule.
For the $f^\prime$ case instead one can consider an intermediate energy
region (say, about 1 GeV for the transferred momentum)
and there the kinetic energy
term could  just provide the ``canonical'' value, since the threshold for the
frozen $\overline{p}$'s is fixed at about two GeV, while the cancellation eq.
(\ref{fpsr3.3}), that has its threshold at quite lower energies,
seems to be reasonably ensured. Thus in the region ranging fro 1 to 2 GeV/c
one could reasonably expect 
\begin{equation}
  \label{eq:last:last}
  \Xi_1^\prime(q)\simeq \left(\frac{1}{2 m}+\sigma\right)A {\bf q}^2
\end{equation}
where $\sigma$ is an {\em a priori} unknown parameter reasonably stable 
in a rather wide energy region.

\vskip2cm

\centerline{\bf Acknowledgements}
Profs. A. Polls and A. Ramos are gratefully acknowledged for the many 
helpful discussions about this topic during my visits at the Dept. of 
Physics of the University of Barcelona. I wish also to thank Prof. G. 
Orlandini for  her  valuable suggestions and advices.
\newpage
%\bibliography{/home/cenni/bibl/references}

\begin{thebibliography}{10}

\bibitem{NoPi-66-B}
{P. Nozi\`eres and D. Pines}.
\newblock {\em The Theory of Quantum Liquids}.
\newblock W. A. Benjamin, inc., New York, Amsterdam, 1966.

\bibitem{OrTr-91}
{G. Orlandini and M. Traini}.
\newblock ~.
\newblock {\em Rept. Prog. Phys.}, 54:257, 1991.

\bibitem{FeWa-71-B}
{A. L. Fetter and J. D. Walecka}.
\newblock {\em Quantum Theory of Many-Particle Systems}.
\newblock McGraw-Hill, New York, 1971.

\bibitem{ItZu-80-B}
{C. Itzykson and J. P. Zuber}.
\newblock {\em {Quantum Field Theory}}.
\newblock McGraw-Hill Book co., Singapore, 1980.

\bibitem{Sc-59}
{J. Schwinger}.
\newblock ~.
\newblock {\em Phys. Rew. Letter}, 3:296, 1959.

\bibitem{Br-66}
{L. S. Brown}.
\newblock ~.
\newblock {\em Phys. Rew.}, 150:1338, 1966.

\bibitem{Am-78}
{D. J. Amit}.
\newblock {\em Field Theory, the Renormalization Group, and Critical
  Phenomena.}
\newblock McGraw Hill, New York, 1978.

\bibitem{Ta-86-B}
Y.~Takahashi.
\newblock {\em Quantum Field Theory}.
\newblock Elsevier Science Publishers, 1986.

\bibitem{LeRuOk-57}
{J. S. Levinger, M. L. Rustgi and K. Okamoto}.
\newblock ~.
\newblock {\em Phys. Rev.}, 106:1191, 1957.

\bibitem{CeDoMo-97}
{R. Cenni, T. W. Donnelly and A. Molinari}.
\newblock ~.
\newblock {\em Phys. Rev.}, C56:276, 1997.

\bibitem{AmCeDoMo-97}
{P. Amore, R. Cenni, T. W. Donnelly and A. Molinari}.
\newblock ~.
\newblock {\em Nucl. Phys.}, A615:353, 1997.

\bibitem{Ba-al-98}
{M. B. Barbaro, R. Cenni, A. De Pace, T.W . Donnelly and A. Molinari}.
\newblock ~.
\newblock {\em Nucl. Phys.}, A643:137, 1998.

\bibitem{AlCeMoSa-88}
{W. M. Alberico, R. Cenni, A. Molinari and P. Saracco}.
\newblock ~.
\newblock {\em Phys. Rev.}, C38:2389, 1988.

\bibitem{Do-al-92}
{T. W. Donnelly et al.}
\newblock ~.
\newblock {\em Nucl. Phys.}, A541:525, 1992.

\bibitem{GoLvPe-83}
{E. I. Gol'braikh, A. I. L'vov and V. A. Petrun'kin}.
\newblock ~.
\newblock {\em Sov. J. Nucl. Phys.}, 37:868, 1983.

\bibitem{Ki-86}
{P. Kienle}.
\newblock ~.
\newblock {\em Ann. Rev. Nucl. Part. Sci.}, 36:605, 1986.

\bibitem{Wa-83}
{J. D. Walecka}.
\newblock ~.
\newblock {\em Nucl. Phys.}, A399:405, 1983.

\end{thebibliography}

\end{document}